\documentclass[12pt,preprint]{aastex}

\shorttitle{HSA Observations of the $z=4.4$ QSO BRI~1335--0417}
\shortauthors{Momjian et al.}

\begin{document}

\title{High Sensitivity Array Observations of the $z = 4.4$ QSO BRI~1335--0417}

\author{Emmanuel Momjian}
\affil{NAIC, Arecibo Observatory, HC 3, Box 53995, Arecibo, PR 00612, USA}
\email{emomjian@naic.edu}

\author{Christopher L. Carilli}
\affil{National Radio Astronomy Observatory, P. O. Box O, Socorro, NM, 87801, USA}
\email{ccarilli@nrao.edu}

\author{Dominik A. Riechers}
\affil{Max-Planck-Institut f\"{u}r Astronomie, K\"{o}nigstuhl 17, D-69117
Heidelberg, Germany}
\email{riechers@mpia.de}

\author{Fabian Walter}
\affil{Max-Planck-Institut f\"{u}r Astronomie, K\"{o}nigstuhl 17, D-69117
Heidelberg, Germany}
\email{walter@mpia.de}

\begin{abstract}

We present sensitive phase-referenced VLBI results on the radio continuum
emission from the $z=4.4$ QSO BRI~1335--0417. The observations were carried
out at 1.4 GHz using the High Sensitivity Array (HSA).
Our sensitive VLBI image at $189 \times 113$~mas ($1.25 \times 0.75$~kpc)
resolution shows continuum emission in BRI~1335--0417 with a total flux 
density of $208 \pm 46~\mu$Jy, consistent with the flux density measured
with the VLA. The size of the source at FWHM is
$255 \times 138$~mas ($1.7 \times
0.9$~kpc) and the derived intrinsic brightness temperature
is $\sim 3.5\times 10^4$~K. No continuum emission is  
detected at the full VLBI resolution ($32 \times 7$ mas, $211 \times 46$~pc),
with a 4$\sigma$
point source upper limit of 34~$\mu$Jy~beam$^{-1}$, or an upper limit to the
intrinsic brightness temperature of $5.6\times 10^5$ K. The highest angular
resolution with at least a 4.5$\sigma$ detection of the radio 
continuum emission is $53 \times 27$~mas ($0.35 \times 0.18$~kpc). At this
resolution, the image shows a continuum feature in BRI~1335--0417
with a size of $64 \times 35$~mas ($0.42 \times 0.23$~kpc) at FWHM,
and intrinsic brightness temperature of $\sim 2\times 10^5$~K.
The extent of the observed continuum sources at 1.4~GHz and the derived
brightness temperatures show that the radio emission (and thus presumably the far-infrared
emission) in BRI~1335--0417 is powered by a major starburst, with a massive star formation
rate of order a few thousand $M_{\odot}~{\rm yr}^{-1}$.  Moreover, the
absence of any compact high-brightness temperature source suggests that there
is no radio-loud AGN in this $z=4.4$ QSO.

\end{abstract}


\section{INTRODUCTION}

Optical surveys such as the Sloan Digital Sky Survey (SDSS; York et al.~2000)
and the Digitized Palomar Sky Survey \citep{DJO99} have revealed large samples
of quasi-stellar objects out to $z \sim 6$. Studies by \citet{FAN02} and \citet{FCK06} have shown that at such a high redshift we are approaching the epoch of
reionization, the edge of the ``dark ages'', when the first stars and massive
black holes were formed.

Observations of high redshift QSOs at mm and sub-mm wavelengths have shown that
a significant fraction ($\sim 30\%$) of the sources are strong emitters of
far-infrared (FIR) radiation, which is thermal emission from warm dust.
These sources have luminosities $L_{\rm FIR} > 10^{12}~L_{\odot}$, and
molecular gas masses greater than $10^{10}~M_{\odot}$ \citep{SV05,RD06}. 
Moreover, the large reservoirs of warm gas and dust in these objects have led to the
hypothesis that these are starburst galaxies with massive star formation rates
on the order of 1000~$M_{\odot}~{\rm yr}^{-1}$ \citep{BER03,BEE06,CAR04,WAL03}.

An important question regarding these high-$z$ QSOs is whether the dust
heating mechanism is dominated by a central AGN, or starburst activity in
the QSO host galaxies. The high resolution of Very Long
Baseline Interferometry (VLBI) observations permits a detailed look at the
physical structures in the most distant cosmic sources. Also, sensitive VLBI
continuum observations can be used to determine the nature of the energy
source(s) in these galaxies at radio frequencies.

To date, several high redshift QSOs have been imaged at milliarcsecond
resolution \citep{FRE97,FRE03,MOM04,BEE04,MOM05}. These are the highest resolution
studies of such distant QSOs by far. In this paper, we present sensitive VLBI
observations of the $z=4.4$ QSO BRI~1335--0417.

The source BRI~1335--0417, which has an optical redshift of $4.396 \pm 0.026$,
was identified in the Automatic Plate Measuring (APM) survey \citep{IMH91,STO96}.
This QSO is the second brightest source at millimeter wavelengths with redshifts
$z > 4$, with the brightest being the $z=4.7$ QSO BRI~1202--0725 \citep{OMO96}.
The implied FIR luminosity of BRI~1335--0417 is
$L_{\rm FIR}=3.1 \times 10^{13}~{L}_{\odot}$. Thus far, there is no evidence for
multiple imaging in BRI~1335--0417 due to strong gravitational lensing \citep{CMY99}. 

The optical spectrum of the QSO BRI~1335--0417 shows a strong absorption by Ly$\alpha$
and metal lines at the source redshift \citep{STO96}. Its CO~$(5-4)$ line emission was detected
at $z=4.4074 \pm 0.0015$ \citep{GUI97}.
The CO~$(2-1)$ emission studies on this source by \citet{CMY99} suggested optically thick emission
from warm ($>$~30~K) molecular gas. The derived total molecular gas mass is $M_{\rm 
H2}=(1.5 \pm 0.3) \times 10^{11}\,M_{\odot}$, making BRI~1335--0417 one 
of the most massive molecular gas reservoirs known at high redshifts.

The source BRI~1335--0417 is also detected in the radio continuum at 1.4 and 4.9~GHz with
the Very Large Array (VLA) in C-configuration, with 
flux densities of $220 \pm 43$ and $76 \pm 11~\mu$Jy, respectively \citep{CMY99,YUN00}.
The source appears unresolved at the resolution of these VLA observations, which is on the order of
several arcseconds.

Throughout this paper, we assume a flat cosmological model with
$\Omega_{m}=0.3$, $\Omega_\Lambda=0.7$, and
${H_{0}=71}$~km~s$^{-1}$~Mpc$^{-1}$. In this model, at the distance of BRI
1335--0417, 1~milliarcsecond (mas) corresponds to 6.6~pc.

\section{OBSERVATIONS AND DATA REDUCTION}

The VLBI observations of BRI~1335--0417 were carried out at 1.4~GHz on 2005 December 31 and
2006 January 7, using the High Sensitivity Array (HSA), which included
the Very Long Baseline Array (VLBA), the phased Very Large Array (VLA), and the
Green Bank Telescope (GBT) of the NRAO\footnote{The National Radio Astronomy
Observatory is a facility of the National Science Foundation operated under
cooperative agreement by Associated Universities, Inc.}. Four adjacent 8~MHz
baseband channel pairs were used in the observations, both with right and
left-hand circular polarizations, and sampled at two bits. The data were
correlated at the VLBA correlator in Socorro, NM, with 2~s correlator
integration time. The total observing time was
14~hr. 
In these observations, the shortest baseline is between the phased VLA and the VLBA antenna in
Pie Town, NM, which is 52~km. This short-spacing limit filters out all spatial structure
larger than about $0\rlap{.}^{''}42$. Table~1 summarizes the parameters of these observations.

The observations employed nodding-style phase referencing, using the calibrator
J1335--0511 ($S_{\rm 1.4~GHz}=0.4$~Jy), with a cycle time of 4~min, 3~min on the
target source and 1~min
on the calibrator. A number of test cycles were also included to monitor the
coherence of the phase referencing. These tests involved switching between two
calibrators, the phase calibrator J1335--0511 and the phase-check calibrator
J1332--0509 ($S_{\rm 1.4~GHz}=0.3$~Jy), using a similar cycle time to that used for
the target source.

The accuracy of the phase calibrator position is important in phase-referencing
observations \citep {WAL99}, as this determines the accuracy of the absolute
position of the target source and any associated components. Phase referencing,
as used here, is known to preserve absolute astrometric positions to better
than $\pm 0\rlap{.}^{''}01$ \citep{FOM99}.

Data reduction and analysis were performed using the Astronomical Image
Processing System (AIPS) and Astronomical Information Processing System
(AIPS$++$) of the NRAO.  After applying {\it a priori} flagging, amplitude
calibration was performed using measurements of the antenna gain and system
temperature for each station. Ionospheric corrections were applied using the
AIPS task ``TECOR''. The phase calibrator J1335--0511 was self-calibrated in
both phase and amplitude and imaged in an iterative cycle.

Images of the phase-check calibrator, J1332--0509, were deconvolved using two
different approaches: (a) by applying the phase and the amplitude
self-calibration solutions of the phase reference source J1335--0511
(Figure~1{\it{a}}), and (b) by self calibrating J1332--0509 itself, in both
phase and amplitude (Figure~1{\it{b}}). The peak surface brightness ratio of
the final images from the two approaches gives a measure of the effect of
residual phase errors after phase referencing (i.e. `the coherence' due to
phase referencing). At all times, the coherence was found to be better than
98\%.

The self-calibration solutions of the phase calibrator, J1335--0511, were
applied on the target source, BRI~1335--0417, which was then deconvolved and imaged at
various spatial resolutions by tapering the visibility data.

\section{RESULTS \& ANALYSIS}

Imaging the target source at the full resolution of the VLBI array, which is 
$32 \times 7$~mas ($211 \times 46$~pc, PA=$-7^{\circ}$), achieved an rms noise level of 
$8.5~\mu$Jy~beam$^{-1}$, but did not reveal any continuum component
in the field of BRI~1335--0417. This indicates the absence of any compact radio
continuum emission with flux densities of $\geq 4\sigma \simeq
34~\mu$Jy~beam$^{-1}$, which in turn implies an upper limit to the intrinsic brightness
temperature (corresponding to a rest frequency of $\sim$8~GHz)
of $5.6 \times 10^5$~K for any compact radio source in
BRI~1335--0417. Our coherence tests during these
observations using two VLBI calibrators show that the lack of a strong point
source in BRI~1335--0417 at the full resolution of the array cannot be due to
the phase referencing procedure. The 4$\sigma$ 
flux limit reported above is almost an order of magnitude lower than the flux measured by 
the VLA ($220 \pm 43~\mu$Jy) at 1.4 GHz \citep{CMY99}.
This immediately implies that more than $\sim$80\% 
of the radio continuum emission in BRI~1335--0417 is extended and not 
confined to the central AGN.

In the following, we will assess if our HSA observations can recover the flux
seen by the VLA. To do so, we applied a two-dimensional Gaussian taper falling to
30\% at 1~M$\lambda$ to the visibility data in both the {\it u}- and {\it v}-directions.
This gives a beam size of $189 \times 113$~mas 
($1.25 \times 0.75$~kpc, P.~A.= $-48^{\circ}$), and the resulting image is shown in Figure~2.
The rms noise level in this naturally weighted image is $28~\mu$Jy~beam$^{-1}$. The
peak flux density of the detected continuum source is $126 \pm 28~\mu$Jy~beam$^{-1}$, and the total
flux density is $208 \pm 46~\mu$Jy, which agrees
well with the flux density measured with the VLA at 1.4~GHz \citep{CMY99}.
The size of the source at full width half maximum (FWHM) is $255 \times 138$~mas 
($1.7 \times 0.9$~kpc), and the derived intrinsic brightness temperature is
$(3.6 \pm 0.9) \times 10^4$~K.

Figure 3 shows the continuum emission in BRI~1335--0417 at the highest angular
resolutions for which there is at least 4.5$\sigma$ detection ($\sigma=19~\mu$Jy~beam$^{-1}$).
This corresponds to  a resolution of $53 \times 27$~mas ($0.35 \times 0.18$~kpc) in
position angle $-6^{\circ}$. This image was obtained by
applying a two-dimensional Gaussian taper falling to 30\% at 5~M$\lambda$
in both the {\it u}- and {\it v}-directions of the visibility
data. The peak flux density of the continuum source detected at this resolution is
$87 \pm 19~\mu$Jy~beam$^{-1}$, and the total flux density is $131 \pm 30~\mu$Jy.
The size of the source is $64 \times 35$~mas ($0.42 \times 0.23$~kpc) at FWHM,
and its intrinsic brightness temperature is
$(2.2 \pm 0.5) \times 10^5$~K. The results obtained at this resolution imply that about
two thirds of the total radio emission emerges from the central $\sim$0.3~kpc of 
BRI~1335--0417, and, likewise, one third arises from more extended scales ($\sim$1.3~kpc).

\section{DISCUSSION}

We have detected 1.4~GHz emission from BRI~1335--00417 (corresponding to a rest frame 
frequency of $\sim$8~GHz) using the HSA. At a moderate resolution ($189 \times 113$~mas; Figure 2)
the intrinsic brightness temperature value of the detected continuum structure is
$\sim 3.5\times 10^4$~K. The measured flux density at this resolution
is consistent with the VLA flux density of this source \citep{CMY99}.
This implies that the radio continuum emission at
1.4~GHz is confined to the extent of the structure seen in the VLBI
image (Figure~2), which is $1.7 \times 0.9$~kpc at FWHM. Furthermore, the highest
resolution image
with continuum detection (Figure~3) shows that almost two thirds of the total
radio continuum emission at 1.4~GHz is arising from the central $\sim$0.3~kpc.

At the full resolution of our array ($32 \times 7$~mas), the radio emission from
BRI~1335--0417 is resolved out and does not show any single dominant source of
very high brightness temperature ($< 5.6\times 10^5$ K). 
This is in contrast to
the results obtained by \citet{MOM04} on a sample of three high-$z$ radio-loud
quasars which were imaged with the VLBA, namely J1053-0016 ($z=4.29$), J1235-0003 ($z=4.69$),
and J0913+5919 ($z=5.11$). In each of these $z>4$ quasars, a
radio-loud AGN dominates the emission at 1.4~GHz on a few mas size scale, with
intrinsic brightness temperatures in excess of $10^9$~K. 

\citet{CON91} have derived an empirical upper limit to the brightness
temperature for nuclear starbursts of $\sim 10^5$~K at 8~GHz (our rest frequency),
while typical
radio-loud AGN have brightness temperatures exceeding this value by at least
two orders of magnitude. These authors also present a possible physical model
for this limit involving a mixed non-thermal and thermal radio emitting (and
absorbing) plasma, constrained by the radio-to-FIR correlation for star-forming
galaxies. 

For BRI~1335--0417, the derived intrinsic brightness
temperatures from our VLBI observations are typical of starburst galaxies.
Also, the linear extent of the radio continuum emission region in
BRI~1335--0417 is typical of local starburst powered Ultra-Luminous IR galaxies
(ULIRGs; Sanders \& Mirabel 1996), such as Arp~220, Mrk~273, and
IRAS~17208--0014 \citep{SMI98,CAR00,MOM03}.

Another argument in favor of a massive starburst origin for the radio continuum
emission from BRI~1335--0417 is that this source follows the radio-FIR correlation for
star forming galaxies, although the massive star formation rate is extreme, of order a few
thousand $M_{\odot}~{\rm yr}^{-1}$ \citep{CMY99}.

The physical characteristics of the QSO BRI~1335--0417 are similar to two other
high-$z$ luminous IR sources; the brightest mm source in the \citet{OMO03} sample of QSOs with
redshifts between $1.8 < z < 2.8$, namely J1409+5628 at $z=2.58$, and the brightest mm
source in the \citet{OMO96}
sample of QSOs at redshifts $z > 4$, namely BRI~1202--0725 at $z=4.7$.
High angular resolution observations of J1409+5628 and BRI~1202--0725 at 1.4~GHz
\citep{BEE04,CAR02} showed
that the radio continuum emission in these optically very bright QSOs is dominated by
extreme nuclear starbursts, with massive star formation rates on the order of
$10^{3}~M_{\odot}~{\rm yr}^{-1}$.  Moreover, sensitive, tapered VLBI images of BRI~1202--0725
at 1.4~GHz \citep{MOM05} revealed the starburst regions in each of its two components that are separated by $4\arcsec$, with
sizes and brightness temperatures comparable to that seen in BRI~1335--0417.

The measured 1.4~GHz radio flux density in BRI~1335--0417 is consistent
with extreme nuclear starburst activity, with the radio continuum being the sum of
supernovae (SNe), supernova remnants, and residual relativistic electrons in the interstellar
medium, as shown in the model presented by \citet{CON91}. However, detecting
individual SNe at $z =4.4$ is unlikely.
\citet{SMI98} reported the detection of several luminous radio SNe in the
prototype ULIRG Arp~220 with flux densities between 0.2 and 1.17 mJy and
angular sizes on the order of 0.25~mas. At the distance of BRI~1335--0417, the
flux densities of such luminous radio SNe would be between $ (0.8-4.5) \times
10^{-3}~\mu$Jy. These values are three to four orders of magnitude lower that the
rms noise levels achieved in our VLBI observations.

\section{ACKNOWLEDGMENTS}

This research has made use of the NASA/IPAC
Extragalactic Database (NED) which is operated by the Jet Propulsion
Laboratory, California Institute of Technology, under contract with the
National Aeronautics and Space Administration. The Arecibo Observatory is part
of the National Astronomy and Ionosphere Center, which is operated by Cornell
University under a cooperative agreement with the National Science Foundation.
C.~L.~C. acknowledges support from the Max-Planck Society
and the Alexander von Humboldt Foundation through the Max-Planck
Forshungspreise 2005.
D.~A.~R. acknowledges support from the Deutsche Forschungsgemeinschaft (DFG)
Priority Programme 1177.

\begin{figure}
\epsscale{0.70}
\plotone{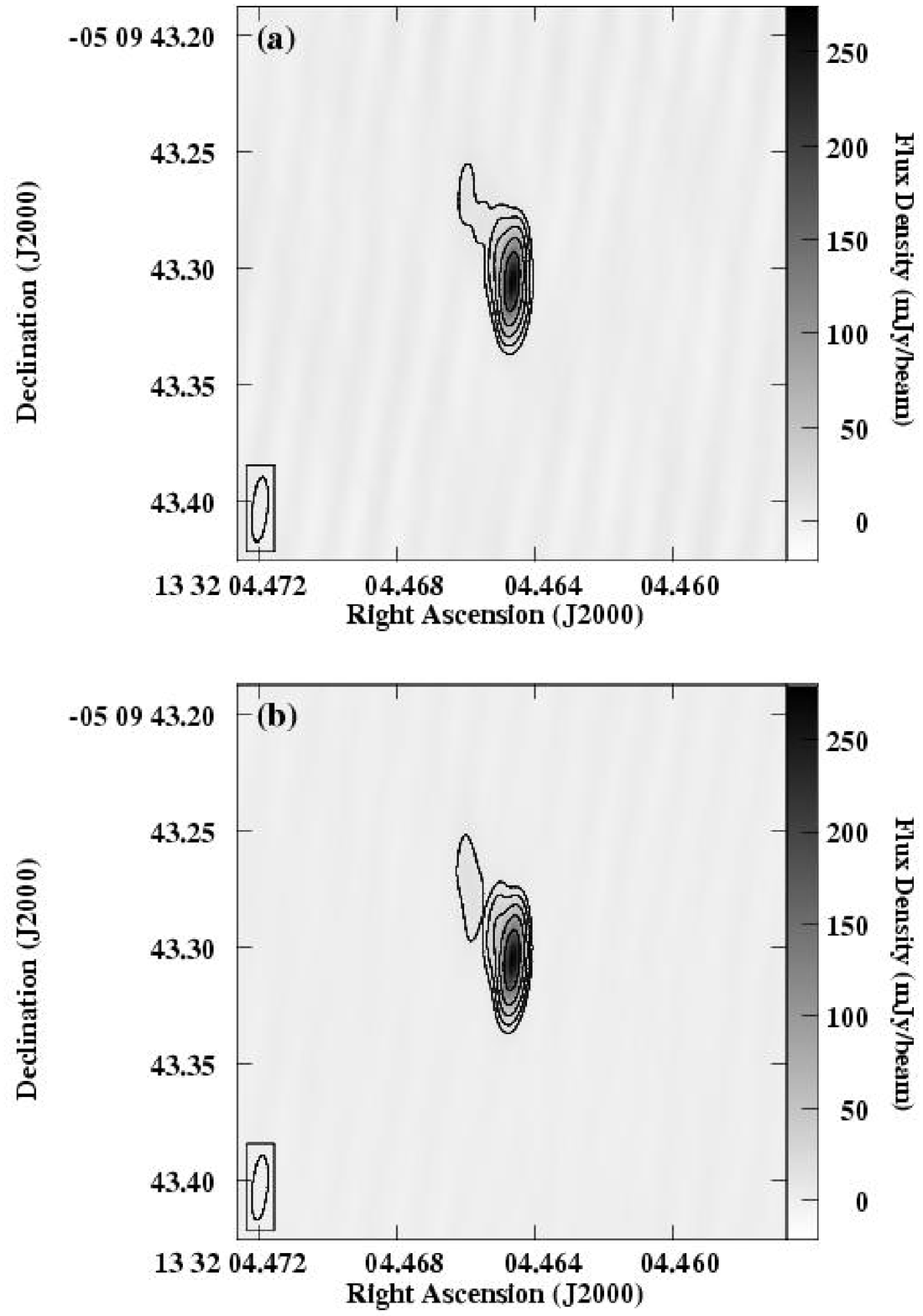}
\figcaption{Continuum images of the phase-check
calibrator J1332--0509 at 1.4~GHz: a) obtained by applying the phase
and the amplitude self-calibration solutions of the phase reference
source J1335--0511, b) obtained by self calibrating J1332--0509
itself, in both phase and amplitude. The restoring beam size in both
images is $27.9 \times 6.5$~mas in position angle $-5^{\circ}$. The
contour levels are at $-3$, 3, 6, 12, 24, 48 times the rms
noise level in the phase-referenced image (top), which is
3.2~mJy~beam$^{-1}$. The gray-scale range is indicated by the step
wedge at the right side of each image.
\label{f1}}
\end{figure}

\clearpage
\begin{figure}
\epsscale{0.8}
\plotone{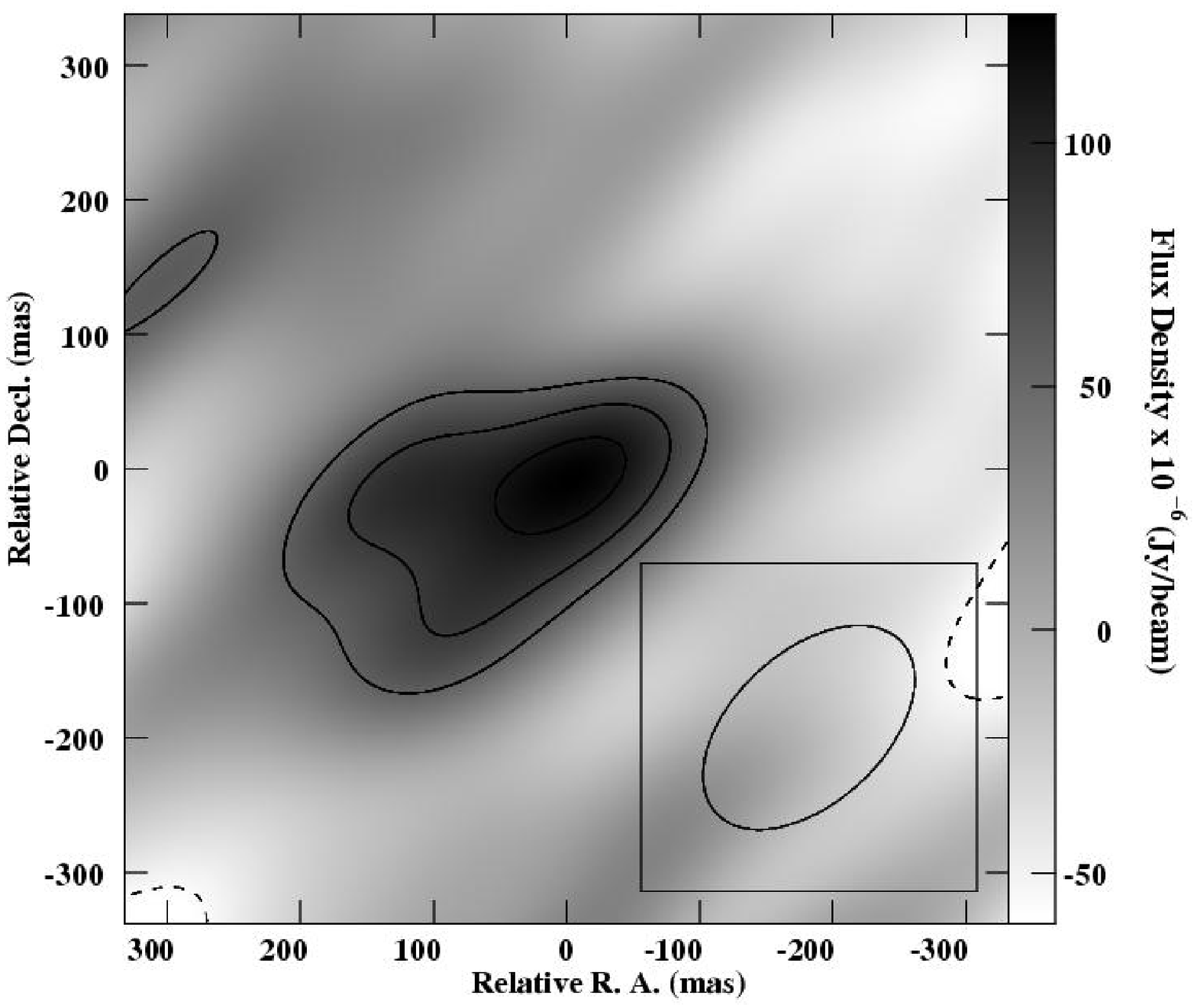}
\figcaption{Naturally weighted 1.4~GHz continuum
image of BRI~1335--0417 at $189 \times
113$~mas resolution (P.~A.=$-48^{\circ}$).  The peak flux density is
126~$\mu$Jy~beam$^{-1}$, and the contour levels are at $-2$, 2, 3, 4
times the rms noise level, which is
28~$\mu$Jy~beam$^{-1}$. The gray-scale range is indicated by the
step wedge at the right side of the image. The reference point (0, 0)
is $\alpha(\rm{J2000.0})= 13^{\rm h}38^{\rm m}03\rlap{.}^{\rm s}4034$,
$\delta(\rm{J2000.0})=-04^{\circ}32^{'}35\rlap{.}^{''}271$.  A two
dimensional Gaussian taper falling to 30\% at 1~M$\lambda$ was
applied.  \label{f2}}
\end{figure}

\clearpage
\begin{figure}
\epsscale{0.8}
\plotone{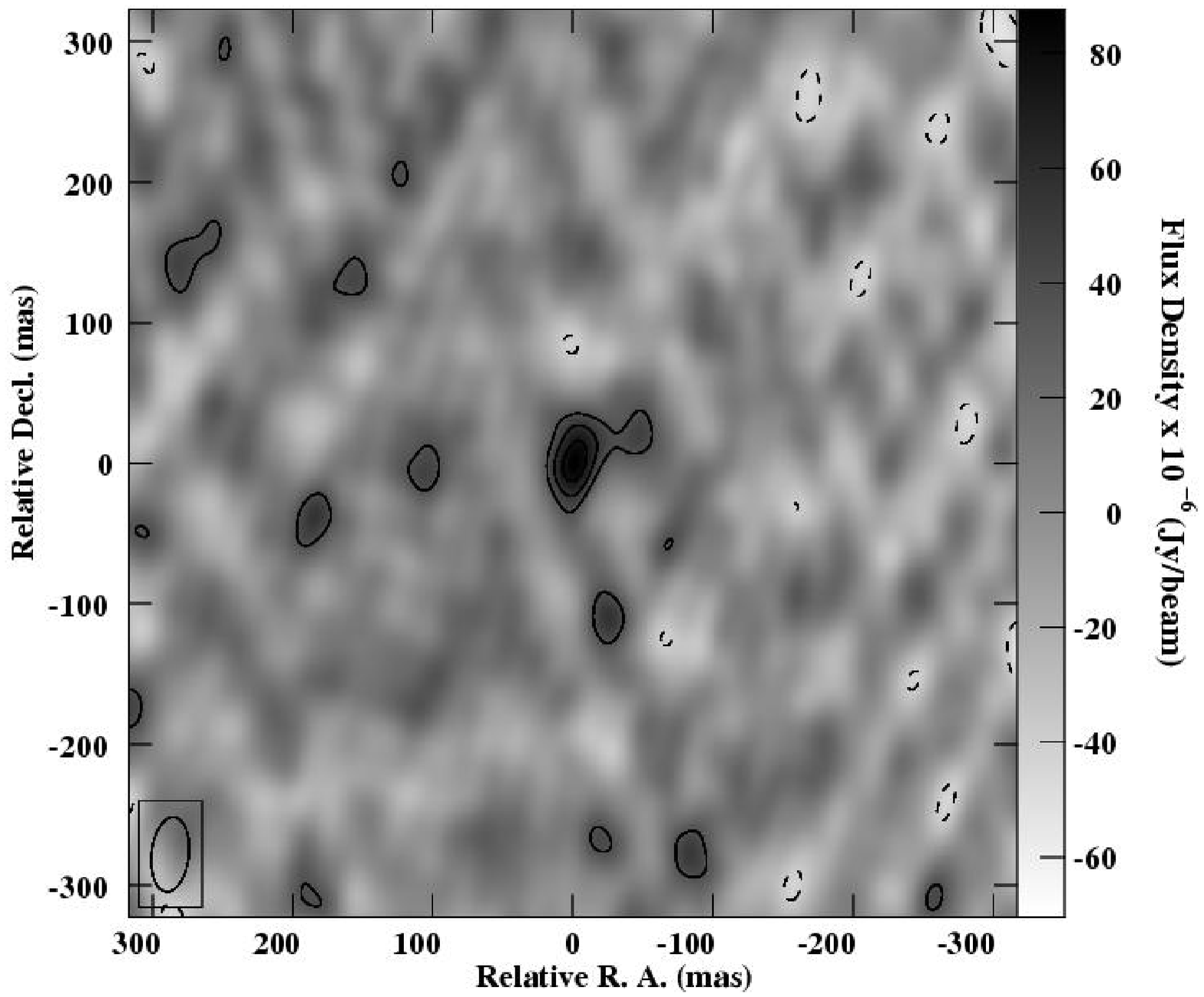}
\figcaption{Naturally weighted 1.4~GHz continuum
image of BRI~1335--0417 at $53 \times
27$~mas resolution (P.~A.=$-6^{\circ}$).  The peak flux density is
87~$\mu$Jy~beam$^{-1}$, and the contour levels are at $-2$, 2, 3, 4
times the rms noise level, which is 19~$\mu$Jy~beam$^{-1}$. The
gray-scale range is indicated by the step wedge at the right side of
the image.  The reference point (0, 0) is $\alpha(\rm{J2000.0})=
13^{\rm h}38^{\rm m}03\rlap{.}^{\rm s}4034$,
$\delta(\rm{J2000.0})=-04^{\circ}32^{'}35\rlap{.}^{''}271$. A two
dimensional Gaussian taper falling to 30\% at 5~M$\lambda$ was
applied.  \label{f3}}
\end{figure}

\clearpage
\oddsidemargin=-1cm
\tabletypesize{\scriptsize}

\begin{deluxetable}{lc}
\tablenum{1}
\tablecolumns{6}\tablewidth{0pc}
\tablecaption{P{\footnotesize ARAMETERS} {\footnotesize OF THE} VLBI
O{\footnotesize BSERVATIONS} {\footnotesize OF} BRI~1335--0417}
\tablehead{\colhead{Parameters} & \colhead{Values}}
\startdata
Observing Dates \dotfill  & 2005 Dec. 31 \& 2006 Jan. 7 \\
Total observing time (hr)\dotfill  & 14 \\
Phase calibrator\dotfill  & J1335--0511 \\
Phase-referencing cycle time (min)\dotfill  &  4 \\
Frequency (GHz)\dotfill  &  1.4 \\
Total bandwidth (MHz)\dotfill   & 32\\
\enddata
\end{deluxetable}

\end{document}